
\magnification 1200
%

%
\font\eightrm=cmr8
\font\eighti=cmmi8
\font\eightsy=cmsy8
\font\eightbf=cmbx8
\font\eighttt=cmtt8
\font\eightit=cmti8
\font\eightsl=cmsl8
\font\sixrm=cmr6
\font\sixi=cmmi6
\font\sixsy=cmsy6
\font\sixbf=cmbx6
\catcode`@11
\newskip\ttglue
\font\grrm=cmbx10 scaled 1200

\def\eightpoint{\def\rm{\fam0\eightrm}
\textfont0=\eightrm \scriptfont0=\sixrm \scriptscriptfont0=\fiverm
\textfont1=\eighti \scriptfont1=\sixi \scriptscriptfont1=\fivei
\textfont2=\eightsy \scriptfont2=\sixsy \scriptscriptfont2=\fivesy
\textfont3=\tenex \scriptfont3=\tenex \scriptscriptfont3=\tenex
\textfont\itfam=\eightit \def\it{\fam\itfam\eightit}
\textfont\slfam=\eightsl \def\sl{\fam\slfam\eightsl}
\textfont\ttfam=\eighttt \def\tt{\fam\ttfam\eighttt}
\textfont\bffam=\eightbf
\scriptfont\bffam=\sixbf
\scriptscriptfont\bffam=\fivebf \def\bf{\fam\bffam\eightbf}
\tt \ttglue=.5em plus.25em minus.15em
\normalbaselineskip=6pt
\setbox\strutbox=\hbox{\vrule height7pt width0pt depth2pt}
\let\sc=\sixrm \let\big=\eightbig \normalbaselines\rm}
\newinsert\footins
\def\newfoot#1{\let\@sf\empty
  \ifhmode\edef\@sf{\spacefactor\the\spacefactor}\fi
  #1\@sf\vfootnote{#1}}
\def\vfootnote#1{\insert\footins\bgroup\eightpoint
  \interlinepenalty\interfootnotelinepenalty
  \splittopskip\ht\strutbox 
  \splitmaxdepth\dp\strutbox \floatingpenalty\@MM
  \leftskip\z@skip \rightskip\z@skip
  \textindent{#1}\footstrut\futurelet\next\fo@t}
\def\fo@t{\ifcat\bgroup\noexpand\next \let\next\f@@t
  \else\let\next\f@t\fi \next}
\def\f@@t{\bgroup\aftergroup\@foot\let\next}
\def\f@t#1{#1\@foot}
\def\@foot{\strut\egroup}
\def\footstrut{\vbox to\splittopskip{}}
\skip\footins=\bigskipamount 
\count\footins=1000 
\dimen\footins=8in 

\def\ref#1{$^{#1}$}
\def\flex{\raise 6pt\hbox{$\leftrightarrow $}\! \! \! \! \! \! }
\def\oversome#1{ \raise 8pt\hbox{$\scriptscriptstyle #1$}\! \! \! \! \! \! }
\def\tr{ \mathop{\rm tr}}

\newbox\bigstrutbox
\setbox\bigstrutbox=\hbox{\vrule height10pt depth5pt width0pt}
\def\bigstrut{\relax\ifmmode\copy\bigstrutbox\else\unhcopy\bigstrutbox\fi}
\def\refer[#1/#2]{ \item{#1} {{#2}} }
\def\rev<#1/#2/#3/#4>{{\it #1\/} {\bf#2}, {#3}({#4})}
\def\boxit#1{\vbox{\hrule\hbox{\vrule\kern3pt
\vbox{\kern3pt#1\kern3pt}\kern3pt\vrule}\hrule}}
\def\gmu{g^{-1}}

\def\sqr#1#2{{\vcenter{\hrule height.#2pt
   \hbox{\vrule width.#2pt height#1pt \kern#1pt
    \vrule width.#2pt}
    \hrule height.#2pt}}}
\def\dal{\mathchoice{\sqr{6}{4}}{\sqr{5}{3}}{\sqr{5}3}{\sqr{4}3} \, }


\def\smin{\,\raise 0.06em \hbox{${\scriptstyle \in}$}\,}
\def\smsubset{\,\raise 0.06em \hbox{${\scriptstyle \subset}$}\,}

\def\Natural{\hbox{\hskip 1.5pt\hbox to 0pt{\hskip -2pt I\hss}N}}

\def\Rational{\hbox{\hbox to 0pt{\hskip 2.7pt \vrule height 6.5pt
                                  depth -0.2pt width 0.8pt \hss}Q}}
\def\Real{\hbox{\hskip 1.5pt\hbox to 0pt{\hskip -2pt I\hss}R}}
\def\Complex{\hbox{\hbox to 0pt{\hskip 2.7pt \vrule height 6.5pt
                                  depth -0.2pt width 0.8pt \hss}C}}
\def \E {{{\rm e}}}
\def \tr {{\rm tr}\, }

\hfill CERN-TH/95-310

\hfill IC/385

\hfill hepth-9511191
\vskip .5cm

\centerline {\grrm  Massive Two-Dimensional Quantum Chromodynamics}

\vskip .6cm

\centerline { E. Abdalla\newfoot {${}^*$}{Permanent address: Instituto de
F\'\i sica - USP, C.P. 20516, S. Paulo, Brazil.} }
\vskip .1cm

\centerline{ International Centre for Theoretical Physics}

\centerline{34.100 Trieste, Italy}
\vskip .3cm

\centerline{M.C.B. Abdalla\newfoot
{${}^\dagger $}{Permanent address: Instituto de F\'\i sica
Te\'orica - UNESP, R. Pamplona 145, 01405-900, S. Paulo, Brazil.}}
\vskip .1cm

\centerline{ CERN-TH}

\centerline{ CH-1211 Geneva 23, Switzerland}
\vskip .1cm
\centerline{and}
\vskip .2cm
\centerline{K.D. Rothe}
\vskip .1cm

\centerline{Institut  f\"ur Theoretische Physik  der Universit\"at Heidelberg}
\centerline{Philosophenweg 16, D-69120 Heidelberg, Germany}

\vskip 1cm

\centerline{\bf Abstract}
\vskip .5cm
\noindent In this work we study the zero-charge sector of massive
two-dimensional Quantum Chromodynamics in the decoupled formulation.
 We find that
some general features of the massless theory, concerning the constraints and
the right- and left-moving character of the corresponding BRST currents,
survive in the massive case. The implications for the integrability
properties previously valid in the massless case, and the structure of the
Hilbert
space are discussed.

\vfill

\noindent CERN-TH/95-310

\noindent IC/385

\noindent hepth/9511191

\vfill\eject

\centerline{\bf 1. Introduction}
\vskip .5cm

The recent formulation of two-dimensional Quantum Chromodynamics
 (QCD$_2$) with massless fermions in terms of positive and negative level
Wess--Zumino--Witten (WZW) fields, ghosts and massive bosonic excitations,
has led to interesting insights  into the characteristics of the model,\ref 1
such
as its integrability,\ref 2 degeneracy of the vacuum,\ref 3 and higher symmetry
algebras related to some operators in the theory.\ref{4,5} Although the fields
of
this equivalent, effective bosonic theory, obtained by making use of the
representation of the fermionic determinant in terms of a
Wess--Zumino--Witten action\ref 6 seem decoupled at the Lagrangian level,
the corresponding sectors are connected by BRST constraints\ref{7,8} operating
on the conformally invariant sector of the theory described by a topological
WZW-coset model.\ref 9 The solution of the corresponding cohomology
problem for $SU(2)$\ref {10} was shown\ref 3 to imply a two-fold degeneracy of
the ground state. The $S$-matrix of the massive sector, expected to describe
the
physical excitations of the theory, factorizes\ref{11} as a consequence of the
infinite number of conservation laws, associated with a particular right-moving
``current" of the model, implying the conservation of individual momenta in
particle scattering amplitudes.

In this paper we extend these investigations to the case where the fermions are
massive. In this case we continue to have two BRST nilpotent charges, as in
massless QCD$_2$, but one of them is modified by the presence of the mass
term. The corresponding BRST currents are again found to be right- and
left-moving.
As expected on general grounds\ref{12} there are first-class constraints
associated
with these currents. They also depend only on the light-cone coordinates $x^-=
(x^0-x^1)$ and $x^+ = (x^0+x^1)$, respectively, and thus are constraints on the
zero-mass sector of the theory. This is consistent with the well-known results
for
the Abelian case,\ref{13} the massive Schwinger model (MSM), where the
longitudinal
part of the fermionic current plays the role of these constraints.\ref{14}

In the MSM, the BRST currents involve the right- and left-moving parts of the
longitudinal current, and the BRST condition on the physical Hilbert space
implies that the positive- and negative-metric interacting bosonic fields can
only
occur in a linear combination corresponding to a zero-mass free field.  This
is again a reflection of the fact that the BRST currents are either right- or
left-moving. In the considerably more complicated non-Abelian case, a similar
situation is found to occur; however, these massless modes are
now given as linear combinations of local operator products of the massive,
group-valued fields in the theory.

As for the integrability  condition obtained in ref. [2] for massless QCD$_2$,
it is
found to be spoiled by the mass term. This is consistent with the
non-integrability
of the MSM, equivalently described by a sine-Gordon theory perturbed by the
mass
term. We conjecture that in the case of massive fermions the  higher
conservation
laws  referred to above are replaced by a constraint on the zero-mass sector.

\vskip 1cm
\penalty -9000
\centerline{\bf 2. BRST currents }
\vskip .5cm
\nobreak
Two-dimensional QCD has several features that are not yet understood. However,
when fermions are integrated over in favour of a bosonic field, several
features
become transparent. In the massless case the procedure has explicitly been
carried
out in ref. [2]. The main steps are summarized in the Appendix. In the
non-local
formulation,\ref 2 the partition function of
two-dimensional QCD with massless fermions is given by\newfoot{\ref *}{As
compared to refs. [1, 2] we drop the tildes in $g$ and $\Sigma$, since no
confusion
arises here. Moreover, $C_-$, $b_{++}$, $b_{--}$, $c_+$ and $c_-$ of refs. [1,
2]
correspond here to $C_-$, $b_+$, $b_-$, $c_-$ and $c_+$, respectively.}
$$
{\cal Z}=\int {\cal D}g {\cal D}\beta {\cal D}\Sigma {\cal D}C\,\E^{iS}\quad,
\eqno(2.1a)
$$
the bosonized action being given in terms of the WZW action, ghosts and
a Yang--Mills term,
$$
S=\Gamma[g] + \Gamma [\beta] - (c_V+1)\Gamma [\Sigma] + S_{gh} +
S_{YM}[\beta, C]  \quad ;\eqno(2.1b)
$$
the Yang--Mills action
$S_{YM}[\beta, C]$,  the ghost  term and the WZW functional $\Gamma[g]$, are
respectively given by
$$
\eqalignno{
S_{YM} &= \int {\rm d}^2x \, \tr \left({1\over 2 }(\partial_+ C)^2
+ \lambda C(\beta^
{-1}i\partial_+\beta)\right)\quad ,&(2.2a)\cr
S_{gh} & = \int {\rm d}^2 x \, \tr \left( c_-i\partial_+b_-
+ c_+ i \partial_-b_+ \right)\quad ,&(2.2b)\cr
\Gamma[g] & = {1\over 8\pi}\int {\rm d}^2x \, \tr \partial_\mu g^{-1}
 \partial^\mu g + {1\over 12\pi}
\int {\rm d}^3 y \epsilon^{\alpha \beta\gamma} \, \tr \, \left(
\tilde g^{-1} \partial_\alpha \tilde g \tilde g^{-1} \partial _\beta \tilde g
\tilde g^{-1} \partial _\gamma \tilde g\right)
\,\, ,&(2.2c)\cr}
$$
with a similar expression for $\Gamma[\beta]$. The parameter $\lambda$ is
given in terms of the charge $e$, and the Casimir $c_V$ by $\lambda =
{c_V+ 1\over 2\pi}e$, where $c_V$ is normalized according to $f^{abc}
f^{dbc}={1\over 2}c_V\delta^{ad}$. The action (2.1) contains the
conformally invariant WZW field  $g$ corresponding to the bosonized version
of the massless free fermionic excitation, $\Sigma$ describes the
negative metric excitations, while the $(\beta , C)$ system corresponds
to the massive sector.  As shown in refs. [2, 7, 8], the
apparently decoupled $g$, $\beta$,  $\Sigma$ and $C$ sectors are actually
connected via BRST constraints.

In the case where the fermions are massive, the functional determinant of the
Dirac operator, an essential ingredient for arriving at the bosonized form
(2.1$a$) of the QCD$_2$ partition function, can no longer be computed in
closed form, and one must resort to the so-called adiabatic principle of
form invariance. Equivalently, one can start with a perturbative expansion
in powers of the mass, as given by
$$
\sum {1\over n!}M^n \left[ \int {\rm d}^2 x\overline \psi \psi\right]^n\quad ,
$$
use the (massless) bosonization formulae and reexponentiate the
result. In this approach, the mass term is given in terms of a
bosonic field $g_\psi $ of the massless theory by\ref{15,16}
$$
S_m= - M\int \overline \psi\psi = M\mu\int \tr (g_\psi + g_\psi^{-1})\quad ,
$$
where $\mu$ is an arbitrary massive parameter whose value depends  on the
renormalization  prescription for the mass operator.\ref{13}

Defining $m^2 = M\mu$, we reexponentiate the mass term. Going through the
changes of variable of ref. [2], one arrives at the following expression in
terms of
the fields of the non-local formulation:
$$
S_m =  m^2 \int\tr (g\Sigma^{-1} \beta + \beta^{-1} \Sigma g^{-1})\quad
.\eqno(2.3)
$$

After such a procedure, the  effective action of massive QCD$_2$ reads
$$
S=\Gamma[g] + \Gamma [\beta] - (c_V+1)\Gamma[\Sigma]+S_{gh}+S_{YM}[\beta, C]+
S_m[g,\beta,\Sigma]\quad , \eqno(2.4)
$$
and the partition function no longer has a factorized form.
Nevertheless, there still exist BRST currents which are either right- or
left-moving.

We wish to construct the BRST currents associated with the above action. This
action exhibits various symmetries of the BRST type; however, not all of them
lead to nilpotent charges. In the case of massless fermions these symmetries
were found to be associated with the transformations

\itemitem{\it a)} \hskip 3 truecm $g\to gY \quad , \quad \Sigma\to \Sigma Y$ ,

\itemitem{\it b)} \hskip 3 truecm $g\to Xg \quad , \quad \Sigma\to X\Sigma$ ,

\itemitem{\it c)} \hskip 3 truecm $\Sigma\to X\Sigma \quad , \quad \beta\to
X\beta$.
\smallskip\noindent
In the massive case, transformation {\it b)} must be supplemented by $\beta\to
X\beta X^{-1}$ in order to leave the mass term invariant.

The respective
BRST-type transformations, leaving the action (2.4) invariant are easily
found. Corresponding to the right transformation of the $g$ and $\Sigma$
fields in item {\it a)} above, we obtain a transformation similar to the
massless case, since such a mapping by itself leaves the mass term invariant:
$$
\eqalignno{
a) \qquad\qquad\delta g =\, & \epsilon gc_+ \quad , \quad\delta\Sigma =
\epsilon \Sigma c_+ \quad ,\cr
\delta C    = \, & 0 \quad , \quad \delta \beta =0 \quad ,\cr
\delta c_- = \, & 0 \quad ,\quad \delta c_+ = {\epsilon \over 2}\{ c_+, c_+\}
\quad ,\cr
\delta b_- = \, & 0 \quad ,\cr
\delta b_+ = \, & \epsilon\left({1\over 4\pi}g^{-1}i\partial _+g -
{c_V +1\over 4\pi}\Sigma^{-1}i\partial_+\Sigma + \{b_+, c_+\}\right)
\quad .&(2.5a)\cr}
$$

Corresponding to the mapping ({\it b}), supplemented by the above-mentioned
transformation $\beta\to X\beta X^{-1}$ in order to leave the
mass term invariant, we obtain
$$
\eqalignno{
b) \qquad\qquad  \delta g  = \, &\epsilon c_- g \quad , \quad \delta \Sigma =
\epsilon  c_-  \Sigma \quad ,\cr
\delta C=\, &\epsilon[c_-, C] \quad , \quad \delta \beta =\epsilon[c_-, \beta]
\quad ,\cr
\delta c_-  = \, & {\epsilon \over 2}\{ c_-, c_-\} \quad ,\quad \delta c_+ =
0\quad ,
\cr
\delta b_+ = \, & 0 \quad ,\cr
\delta b_- = \, & \epsilon \Biggl( {1\over 4\pi}g i\partial _- g^{-1} -
{c_V +1\over
4\pi}\Sigma i\partial_-\Sigma^{-1} + \{b_-, c_-\}\Biggr)+ \epsilon {\cal B}
\quad,
&(2.5b)\cr}
$$
where
$$
{\cal B} = {1\over 4\pi}\beta i\partial_-\beta^{-1} + {1\over
4\pi\lambda}\partial_+
\partial_-C -\lambda [\beta, C\beta^{-1}]+i[C, \partial_+C].\eqno(2.6)
$$

 The term $\epsilon \cal B$ arises from the
transformation $\beta \to X\beta X^{-1}$, which is unnecessary in the massless
case, and which leads to the transformation law (2.5$b$)
coupling the Yang-Mills sector of the model to the remaining sectors.
Finally, corresponding to the third transformation {\it c)} we have
$$
\eqalignno{
c) \qquad\qquad \delta g = \, & 0 \quad , \quad \delta \Sigma = \epsilon  c_-
\Sigma \quad ,\cr
\delta C = \, & 0 \quad , \quad \delta \beta = \epsilon c_-\beta \quad ,\cr
\delta c_-  =\, & {\epsilon\over 2}\{c_-, c_-\}\quad ,\quad \delta c_+ = 0
\quad ,\cr
\delta b_+ = \, & 0 \quad ,\cr
\delta b_- = \, & \epsilon\left( {1\over 4\pi}\beta i\partial _-
\beta^{-1} - {c_V +1\over 4\pi}\Sigma i\partial_-\Sigma^{-1} - \lambda \beta C
\beta^{-1} +  \{b_-, c_-\}\right)\,\,.&(2.5c)\cr}
$$
This symmetry transformation is again analogous to the one
found in the massless case.

The equations of motion are obtained from action (2.4) by computing its
variation. We obtain
$$
\eqalignno{
{1\over 4\pi}\partial_+(g\partial_-g^{-1}) =\, &  m^2 (g\Sigma^{-1}\beta -
\beta^{-1} \Sigma g^{-1})\quad ,&(2.7a)\cr
-{c_V+1\over 4\pi}\partial_+(\Sigma \partial_-\Sigma^{-1}) =
\, & m^2 (\Sigma g^
{-1}\beta^{-1} - \beta g \Sigma^{-1})\quad, &(2.7b)\cr
{1\over 4\pi} \partial_+ (\beta \partial_-\beta^{-1}) +
i \lambda\partial _+(\beta C
\beta^{-1}) =
\, & m^2 (\beta g \Sigma^{-1} - \Sigma g^{-1}\beta^{-1})\quad, &(2.7c)\cr
-{1\over 4\pi}\partial_-(\beta^{-1} \partial_+ \beta) + i\lambda [\beta^{-1}
\partial_+ \beta, C] + i \lambda \partial_+C = \, & m^2 (g\Sigma^{-1}\beta -
\beta ^{-1}\Sigma g^{-1})\quad ,&(2.7d)\cr
\partial_+^2 C = \, & \lambda (\beta ^{-1}i\partial_+\beta)\quad ,&(2.7e)\cr
\partial_\pm b_\mp = \, & 0 \quad ,
\quad \partial _\pm c_\mp =0 \quad .&(2.7f)\cr}
$$
Notice the form of the mass term, which can be transformed, from one equation
to another, by a suitable conjugation. Making use of eqs. (2.7), the Noether
currents are constructed in the standard fashion. The only subtlety in this
procedure concerns the WZW term, which only contributes off shell to the
variation. The three conserved Noether currents are  found to be
$$
\eqalignno{
J_+^{(1)} = \, &\tr c_+ \left({1\over 4\pi}g^{-1}i\partial _+g -
{c_V +1\over 4\pi}
\Sigma^{-1}i\partial_+\Sigma +{1\over 2} \{b_+, c_+\}\right)\quad,
 &(2.8a)\cr
J_-^{(2)} = \, &\tr c_- \left( {1\over 4\pi}g i\partial _- g^{-1} -
{c_V +1\over 4\pi}
\Sigma i\partial_-\Sigma^{-1} + {1\over 2}\{b_-, c_-\} + \cal B \right)\quad,
 &(2.8b)\cr
J_-^{(3)} =\, & \tr c_- \left( {1\over 4\pi}\beta i\partial _- \beta^{-1} -
 {c_V+1\over 4\pi} \Sigma i\partial_-\Sigma^{-1} -
 \lambda \beta C \beta^{-1} + {1\over 2}
\{b_-, c_-\}\right) \quad , &(2.8c)\cr}
$$
and are either ``right" or ``left" moving, that is,
the equations of motion read
$$
\partial_-J_+^{(1)}=0 \quad ,\quad \partial _+J_-^{(2)}=0 \quad ,\quad
\partial_+J_-^{(3)}=0 \quad .\eqno(2.9)
$$

It is convenient to write these currents in the form
$$
\eqalignno{
J_+^{(1)}=\, & \tr \left( c_+ \Omega^{(1)} -
{1\over 2} b_+ \{c_+, c_+ \}\right)
\quad ,&(2.10a)\cr
J_-^{(r)}=\, & \tr \left( c_- \Omega^{(r)} -
{1\over 2} b_- \{c_-, c_- \}\right)
\quad , \quad r= 2,3 \quad ,&(2.10b)\cr}
$$
where $\Omega^{(r)}$ are seen to be given by
$$
\eqalignno{
\Omega^{(1)} = \, & {1\over 4\pi}g^{-1}i\partial _+g - {c_V +1\over 4\pi}
\Sigma^{-1}i\partial_+\Sigma + \{b_+, c_+\}  \quad,     &(2.11a)\cr
\Omega^{(2)} = \, & {1\over 4\pi}g i\partial _- g^{-1} - {c_V +1\over 4\pi}
\Sigma i\partial_-\Sigma^{-1} + \{b_-, c_-\} + \cal B \quad , &(2.11b)\cr
\Omega^{(3)} =\, & {1\over 4\pi}\beta i\partial _- \beta^{-1}
- {c_V+1\over 4\pi}
\Sigma i\partial_-\Sigma^{-1} - \lambda \beta C \beta^{-1} +  \{b_-, c_-\}
 \quad . &(2.11c)\cr}
$$

These operators obey simple equations as a consequence of the current
conservation equations, namely $\Omega^{(1)}$ is right-moving while $\Omega^
{(2,3)}$ are left-moving. Indeed, making use of the equation of
motion (2.7) one
readily checks that the operators $\Omega^{(1)}$, $\Omega^{(2)}$, and
$\Omega^{(3)}$ satisfy
$$
\partial_-\Omega^{(1)}=0 \quad ,\quad \partial _+\Omega ^{(2)}=0 \quad ,\quad
\partial_+\Omega ^{(3)}=0 \quad ,\eqno(2.12)
$$
consistent with the conservation laws (2.9).

In order that the corresponding charges $Q^{(r)}$ be nilpotent, the operators
$\Omega^{(r)}$ should be first class. We examine this question in the following
section.

\vskip 1cm
\penalty-3000
\centerline {\bf 3.  Anomalous constraints, physical subspace}
\vskip .5cm
\nobreak

To establish the first- and second-class character of the operators $\Omega^{
(r)}\,,\,r=1,2,3$,  we rewrite the operators (2.11) in terms of canonical
phase-space
variables. Following the canonical quantization procedure of ref. [17], we have
$$
\eqalignno{
\Omega^{(1)}= \, & -i\hat \Pi^g g + {i\over 4\pi} g^{-1} g' - i\hat \Pi^\Sigma
\Sigma - i {c_V+1\over 4\pi} \Sigma^{-1}\Sigma'+ j_+^{gh} \quad ,&(3.1a)\cr
\Omega^{(2)} = \, & i g \hat \Pi^g  + {i\over 4\pi} g' g^{-1} +i\Sigma\hat\Pi^
\Sigma - i {c_V+1\over 4\pi} \Sigma'\Sigma^{-1}+ j_-^{gh} \cr
& + i\beta \hat\Pi^\beta + {i\over 4\pi} \beta'\beta^{-1} -i\hat
\Pi^\beta \beta +  {i\over 4\pi} \beta^{-1}\beta' -
{1\over 2\pi \lambda} \Pi'_C + i[C, \Pi_C]\quad ,
&(3.1b)\cr
\Omega^{(3)} = \, &  i\beta \hat\Pi^\beta + {i\over 4\pi} \beta'\beta^{-1}  +i
\Sigma\hat\Pi^\Sigma  -i{c_V+1\over 4\pi} \Sigma'\Sigma^{-1} + j_-^{gh}\quad .
&(3.1c)\cr}
$$
where
$$
j^{gh}_{\pm}= \{b_{\pm},c_{\pm}\}.\eqno(3.1d)
$$
Although in the bosonized formulation quantum anomalies arising from one-loop
fermion graphs are already incorporated on the semi-classical  level, the
commutators of the operators $\Omega^{(r)}$ may still be non-canonical\ref{18}
due to the presence of other types of anomalies.\newfoot{$^*$}{In the absence
of
further anomalies these commutators are identical with $i$ times their
respective
Poisson brackets.} For the operators $\Omega^{(1)}$ and $\Omega^{(3)}$ this
situation does not occur. Therefore we are able in this case to compute  their
Poisson brackets using the canonical formalism.  Concerning $\Omega^{(1)}$, it
is
straightforward  to verify that it obeys a Kac--Moody algebra with vanishing
central
term, being  therefore first class with respect to itself. Moreover, since the
Poisson
brackets respect chirality, it also has vanishing Poisson brackets with the
other
operators, $\Omega ^{(2)}$ and  $\Omega^{(3)}$. The  commutator of
$\Omega^{(3)}$
with itself is given in terms of the known Poisson brackets of Kac--Moody
currents,
and id found to vanish weakly. Hence $\Omega^{(3)}$ is also first class with
respect to
itself, and moreover it commutes with $\Omega^{(1)}$. Summarizing, we have,
$$
\eqalignno{
[\Omega^{(1)a}(x), \Omega^{(1)b}(y)] = \, & i f^{abc}\Omega^{(1)c}(x)
\delta(x-y)
\quad, &(3.2a)\cr
[\Omega^{(3)a}(x), \Omega^{(3)b}(y)] = \, &  if^{abc}\Omega^{(3)c}(x)
\delta(x-y)
\quad, &(3.2b)\cr
[\Omega^{(1)a}(x), \Omega^{(3)b}(y)] = \, & 0 \quad . &(3.2c)\cr}
$$

 As for $\Omega^{(2)}$ the situation is
more delicate. It is convenient to write $\Omega^{(2)}$ in the form
$$
\Omega^{(2)} =\Omega^{(3)} + \widetilde\Omega^{(2)}\quad ,\eqno(3.3a)
$$
where now
$$
\widetilde\Omega^{(2)}= {1\over 4\pi \lambda}\partial_+\partial_- C+
\lambda C +
i[C, \partial_+C] + {i\over 4\pi} g \partial_- g^{-1}\quad .\eqno(3.3b)
$$
Our discussion can be restricted to $\widetilde\Omega^{(2)}$, which is also
right-moving. Unlike the previous case, the computation of the Poisson bracket
of this operator with itself involves quantum corrections,
arising from the presence of
the algebraic commutator $[C,\Pi_C]$, which contribute to the central charge.
These quantum corrections are obtained via the
short-distance expansion
$$
[C(x),\Pi_C(x)]_{ij}[C(y),\Pi_C(y)]_{kl} ={2N\over (x-y)^2}\left(
\delta_{il}\delta_{kj}-{1\over N}\delta_{ij}\delta_{kl}\right)\; ,\eqno(3.4)
$$
valid for a symmetry group $G=SU(N)$. Thus we arrive at the result
$$
[{\widetilde\Omega^{(2)a}}(x),{\widetilde\Omega^{(2)b}}(y)]=
i f^{abc}{\widetilde\Omega^{(2)c}}(x)\delta (x-y) + \left[{N\over\pi}\right]
\delta'(x-y)\delta^{ab}\quad .\eqno(3.5)
$$
Similarly
$$
[\Omega^{(2)a}(x), \Omega^{(2)b}(y)] = i f^{abc}{\Omega^{(2)c}}(x)\delta (x-y)
+
\left[{2N+1\over 2\pi}\right]\delta'(x-y)\delta^{ab} \quad .\eqno(3.6)
$$
Therefore these operators are second class. This is analogous to the case of
QCD$_2$ with massless quarks. Thus only the charges associated with $J^{(1)}$
and $J^{(3)}$ are nilpotent and lead to bona fide BRST charges.

Using the Karabali--Schnitzer (KS) method we find that $\Omega^{(1)}$ and
$\Omega^{(3)}$ are constrained to vanish. This is also consistent with the
BRST conditions, which require that the
physical states be annihilated by the BRST charges:

$$
Q^{(1)} \vert {\rm Phys}\rangle =0\; , \quad {\rm and}\quad  Q^{(3)}\vert {\rm
Phys}
\rangle =0
\quad .\eqno(3.7)
$$
In the massless case the solution of the corresponding cohomology problem
only involves the WZW fields
 $g$, $\Sigma$, and then ghosts, being equivalent to a
$G/G$ topological field theory, for the vacuum sector. This problem has been
explicitly solved for the gauge group $SU(2)$ using the Wakimoto
representation for the currents,\ref{19} and
the representation theory of affine algebras.\ref{20}
In the massive case the problem is more
involved, since the partition function no longer
factorizes.
\vskip1cm
\penalty-3000
\centerline{\bf 4. Abelian case}
\vskip .5cm
\nobreak

It is curious that there exists no KS gauging\ref{7} of the action (2.1$b$),
which
would establish $\widetilde\Omega^{(2)}\approx 0$ as one further constraint.
Of course, since $\tilde\Omega^{(2)}$ is second class with respect to itself,
the
associated charge $\widetilde Q^{(2)}$ is not nilpotent, and there is no
compelling reason for $\widetilde\Omega^{(2)}$ to be constrained to vanish.
Nevertheless, in the massless case there does exist one further constraint. In
order to gain some insight into this curious fact, it is instructive to
specialize the
above results to the Abelian case (massive Schwinger  model (MSM)),  where a
plethora of results are available in the literature.\ref{13}

In order to allow for a simple comparison with the standard results on the
massive Schwinger model, we  parametrize the non-Abelian fields as follows
$$
g= \E^{i2\sqrt\pi\varphi}\; ,\quad \tilde\Sigma=\E^{-i2\sqrt\pi\eta}\; ,\quad
\beta=\E^{-i2\sqrt\pi E}\quad .\eqno(4.1)
$$

The equations of motion (2.7) then reduce to $(\lambda = e/2\pi)$
$$
\eqalignno{
\dal \varphi = -\dal\eta =\, & 4\sqrt \pi m^2 \sin 2 \sqrt \pi (E -\varphi -
\eta)\quad , &(4.2a,b)\cr
\dal E + {e\over \sqrt\pi} \partial_+ C =\, & - 4\sqrt\pi m^2 \sin 2 \sqrt \pi
(E - \varphi - \eta)\quad , &(4.2c,d)\cr
\partial_+\left(\partial_+ C - {e\over \sqrt \pi  }E\right) = \, & 0 \quad .
&(4.2e)\cr}
$$

The constraints $\Omega^{(1)}\approx 0$ and $\Omega^{(3)}\approx 0$  read
in this case,
$$
\eqalignno{
2\sqrt \pi \Omega^{(1)} = \,  & -\partial _+(\varphi + \eta ) = -\left[ (\Pi_
\varphi + \partial_1\varphi)+(-\Pi_\eta + \partial_1\eta)\right] \approx 0
\quad ,& (4.3a)\cr
2\sqrt \pi \Omega^{(3)} = \,  & - \partial _- (E- \eta )  - {e\over \sqrt
\pi } C = - \left[ (\Pi_E - \partial_1E)+(\Pi_\eta + \partial _1\eta)\right]
\approx 0 \quad ,
& (4.3c)\cr}
$$
where the canonical momenta are given by the expressions
$$
\Pi_\varphi = \partial_0\varphi \quad ,\quad \Pi_\eta = - \partial_0\eta\quad ,
\quad\Pi_C=\partial_+ C\quad, \quad \Pi_E= \partial_0 E
+{e\over {2\sqrt\pi}}C \quad .
\eqno(4.4)
$$

The constraint  $\Omega^{(1)} \approx 0 $ defines the physical Hilbert space in
the MSM corresponding to positive  chirality.\ref 3  For $\widetilde
\Omega^{(2)}$
one obtains
$$
\eqalignno{
2\sqrt\pi\widetilde\Omega^{(2)}= \, & \partial_- \varphi + {\sqrt\pi\over e}
\partial_+ \partial_- C + {e\over \sqrt\pi} C \cr
= \, & (\Pi_\varphi - \partial_1\varphi) + (\Pi_E - \partial_1 E) +
2\partial_1\left(E -{\sqrt \pi\over e}\Pi_C\right) \quad. &(4.5)\cr}
$$
Making use of (4.3$c$), and supposing that the equation of motion (4.2$e$) has
only the trivial solution $\Pi_C = {e\over \sqrt\pi}E$ (i.e. assuming the
operator
$\partial_+$ to be ``invertible''),  $\widetilde\Omega^{(2)}$ reduces to
${1\over 2
\sqrt \pi}\partial_-(\varphi +\eta) $, which, following the method of
Karabali--Schnitzer,\ref{7} is easily  shown to be constrained to vanish. Thus
$\partial_-\widetilde\Omega^{(2)} = 0$ is replaced by the constraint
$\widetilde
\Omega^{(2)} = 0$. As is well known,\ref{3,13} the constraint $\partial_-
(\varphi +
 \eta) = 0$ defines the Hilbert space of the MSM corresponding to negative
chirality. The constraints $\partial_\pm(\varphi +\eta) \approx 0 $
are indeed (first-class) constraints of the MSM.

The ``invertibility requirement" referred to above can be appreciated by
formally rewriting the constraint (4.3$c$) in configuration space as:
$$
-2\sqrt\pi \Omega^{(3)} = \partial_+^{-1}\left\{ \left(\dal E + {e\over
\sqrt\pi}\right) - \dal \eta \right\} \approx 0 \quad.\eqno(4.6)
$$

Using the equation of motion (4.2) we see that the constraint is guaranteed by
the equations of motion, provided the operator $\partial_+$ is invertible. This
corresponds to the requirement that massless states in the combination (4.6)
be absent.

\vskip 1cm
\penalty-3000
\centerline{\bf 5. The fate of the integrability condition}
\vskip.5cm
\nobreak
The operator (3.3$b$) may be written in the form
$$
\widetilde \Omega^{(2)} = I_-+ {1\over 4\pi } g i \partial_- g^{-1}\quad ,
\eqno(5.1)
$$
where $I_-$ is the operator
$$
I_-= {1\over 4\pi \lambda }\partial_+\partial_- C +  i [C, \partial_+C] +
\lambda C
\quad, \eqno(5.2)
$$
introduced in ref. [2]. Making use of the equation of motion (2.7$a$), we have
$$
\partial_+I_-=-m^2(g\Sigma^{-1}\beta-\beta^{-1}\Sigma g^{-1} )\quad .\eqno(5.3)
$$

In the massless case, $\partial_+I_- = 0$, so that $I_-$ is
right-moving. In refs. [2,11] this was shown to
imply the existence  of an infinite number of charges $Q^{(n)}$, acting on
asymptotic states $\vert \vec p\rangle$ in the massive $\beta$-sector like
generators $Q^a_R$ of $SU(N)_R$, with multiplication by the nth power of the
momenta $p_-$:
$$
Q^{a(n)}\vert \vec p\rangle = p_-^nQ^a_R\vert \vec p\rangle \quad .\eqno(5.4)
$$

It has further been shown in [11] that the conservation laws implied by (5.4)
determine the $S$-matrix in the $\beta$ sector, up to bound-states poles.
Further results implying integrability of the model were also found by
other authors.\ref{21}

For $m\ne 0$ it is evident from (5.3) that the conservation law $\partial_+
I_-=0$ of the massless theory is being replaced by $\partial_+\widetilde
\Omega^{(2)}=0$. This property is evidently guaranteed by the equation of
motion (2.7$c$). We could require $\widetilde \Omega^{(2)}$ itself to vanish.
 As we illustrated in the preceding section for the case of the MSM,
this requirement is easily seen from (2.7$c$) to reflect the absence of zero
modes of the operator $\partial_+$, and is thus a condition on the zero-mass
sector of the theory, rather than an integrability condition. Of course we
could require such a
constraint also in the massless case. However, in that case we have in each
sector two conservation laws, for $I_-$ and $gi\partial_-g^{-1}$  separately!

In principle, the operator $\widetilde\Omega^{(2)}$ is not constrained to
vanish in the general massive non-Abelian case. However, it is clear
that it is a massless field. Even more, it contains only right-moving
excitations. We can speculate whether such excitations are remnants
of colour states. In that case, it would be rather desirable to require such
an operator to vanish, in which case the fields constrained in this way would
be free
from such zero-mass colour excitations, as a consequence of this requirement.
On the other hand, the operator $I_-$, which was connected, in the massless
case,
to an infinite number of conservation laws, is now no longer conserved on its
own;
instead, its conservation is spoiled by the mass term, and
the previous conclusions for the integrability of the $\beta$ sector
can no longer be drawn. Instead, a relation to the $g$ sector is
established.

\vskip 1cm
\penalty-3000
\centerline{\bf 6. Conclusions}
\vskip .5cm
\nobreak

In this paper we have extended the analysis of massless QCD$_2$ from  refs.
[2, 8] to the case where the fermions are massive. In particular we have
obtained the BRST currents in the non-local formulation of ref. [2], thereby
extending the BRST analysis of ref. [8] to the massive case. It is interesting
that these currents again turned out to be either right- or left-moving. As a
result, the BRST condition implied restrictions on the massless sector of the
physical Hilbert space  in particular the existence of two first-class
constraints depending only on one of the light-cone coordinates $x^\pm$. These
first-class constraints were also obtained by appropriately gauging the
partition function, following the method of Karabali and Schnitzer.\ref{7}

In the massless case the solution of the associated cohomology problem for
$SU(2)$ revealed a two-fold degeneracy of the right- and left-handed vacuum
sector. In that case the physical states could be characterized in each
sector by the heighest weight eigenstates of the third component of the
isospin of two conformal WZW currents and of the ghost current. In the massive
case, the situation remains the same for the ghost sector, but we are left
with only one combination of those WZW currents, serving to label the states.
As a result we expect the ground state to no longer be degenerate. This would
be in accordance with the  well-known results on the massive Schwinger model.
The implications of this for the vacuum structure of massive QCD$_2$ is
presently under investigation.

Finally, problems related to the description of physical properties of the
model concerning chiral symmetry breaking\ref{22} as well as the issue of
screening versus confinement,\ref{23} can be analysed.

\vskip 1cm

\centerline {\bf Appendix}
\vskip .5cm

The implementation of the bosonization techniques of a non-Abelian symmetry
is well known.\ref{} In two dimensions we can locally
write the gauge field in terms of two matrix-valued fields $U$ and $V$ as
$$
A_+ = {i\over e}U^{-1}\partial _+U \quad , \quad
A_- = {i\over e}V\partial _-V^{-1} \quad . \eqno(A1)
$$

The effective action $W[A]$ is obtained by integrating the functional
differential
equations associated with the conservation of the vector current, and the
anomaly
in the axial vector currrent. One finds $ W[A]=  - \Gamma [UV] $, where $\Gamma
[UV] $ is the Wess--Zumino--Witten (WZW) functional, obeying the
Polyakov--Wiegmann identity
$$
\Gamma[UV]=\Gamma[U]  +\Gamma[V]+{1\over 4\pi}\tr\,\int{\rm d}^2x\,
U^{-1}\partial_+UV \partial_- V^{-1} \quad .\eqno(A2)
$$

In order to implement the change of variables ($A$1), in the quantum theory,
we still have to compute its Jacobian, that is
$$
J=\det{\delta A_+\over\delta U}{\delta A_-\over\delta V}=\det {\nabla} = \E^
{-i c_V \Gamma[UV]}\quad .\eqno(A3)
$$

It is well known that the invariances of the fermionic part of the QCD
Lagrangian  under local $SU(N)$, as well as $SU(N)_L\times SU(N)_R$
transformations,  $U\to U w^{-1}$ and $V\to w^{-1}V$, are not symmetries of the
effective action $W[A]$ due to the axial anomaly. As a consequence we find
$$
\det i\not\!\!D\equiv\E^{iW[A]}=\int{\cal D}g\,\E^{iS_F[A,g]}\quad,\eqno(A4)
$$
where $S_F(A,g)$ plays the role of an equivalent bosonic action
$$\eqalign{
S_F[A,g]&=\Gamma[g]+{1\over 4\pi}\!\int\!{\rm d}^2x\,{\rm tr}\, \left[ e^2A^\mu
A_ \mu
-e^2A_+gA_-\gmu -eiA_+g\partial_-\gmu -eiA_-\gmu \partial _+g\right]\cr
&=\Gamma[UgV]-\Gamma[UV]\quad.\cr}\eqno(A5)
$$

Using ($A$5), we have for the partition function
$$
{\cal Z} = \int {\cal D}g {\cal D}A_\mu \, \E ^{i \int {\rm d}^2 z \, \{ S_F
[A,g] -
{1\over 4} \tr F_{\mu\nu}F^{\mu\nu} \} }\quad .\eqno(A6)
$$

Using further the identity
$$
\E^{-{i\over 4}\int{\rm d}^2z\,\tr F_{\mu\nu}F^{\mu\nu}}=\int{\cal D}E\,
\E^{-i\int {\rm d}^2z\,\left[{1\over 2}\tr E^2+{1\over 2}\tr EF_{+-}\right]}
\quad,\eqno(A7)
$$
where $E$ is a matrix-valued field, we may rewrite ($A$6) as
$$
{\cal Z}= \int {\cal D}E{\cal D} U{\cal D}V{\cal D}g  \E^{i\Gamma[UgV]
-i(c_V+1)
\Gamma[UV] -i \int {\rm d}^2 z \,\tr [{1\over 2} E^2 + {1\over 2}E F_{+-}] }
\quad . \eqno(A8)
$$
Defining a new gauge-invariant field $\widetilde g=UgV$, and using the
invariance
of the Haar measure, ${\cal D}g={\cal D}\widetilde g$, we see that the field
$\widetilde g$ decouples in the partition function:
$$
{\cal Z}=\int{\cal D}\widetilde g\,\E^{i\Gamma[\tilde g]}\int{\cal D}E{\cal D}U
{\cal D} V{\cal D}({\rm ghosts})\E^{-i(c_V+1)\Gamma[UV]-i\int{\rm d}^2z\, \tr[
{1\over 2}E^2+{1\over 2}EF_{+-}]+iS_{\rm ghosts}}.\eqno(A9)
$$
Introducing the new variable $\Sigma = UV$, we have the identity
$$
\tr EF_{+-}={i\over e}\tr UEU^{-1}\partial_+(\Sigma\partial_-\Sigma^{-1})
\quad .\eqno(A10)
$$
It is natural to redefine variables as $\widetilde E \approx UEU^{-1}, {\cal D}
E =
{\cal D}\widetilde E$, where we have used again the invariance of the Haar
measure.
The partition function ($A$9) then reduces to the form (we choose the gauge
$U=1$)
$$
{\cal Z}\!=\!\!\int\!{\cal D}\widetilde g\,\E^{i \Gamma[\tilde g]}{\cal D}({\rm
gh})\, \E^
{iS_{{\rm gh}}}\! {\cal D} \Sigma{\cal D}\widetilde E\E^{-i(c_V+1)\Gamma [
\Sigma]-
(c_V+1)\tr\!\int \!{\rm d}^2z\partial_+\widetilde
E\Sigma\partial_-\Sigma^{^{-1}}\!-
\!2ie^2 (c_V+1)^2 \!\int \!{\rm d}^2z\tr \widetilde E^2 }.\eqno(A11)
$$
In order to arrive at the partition function (2.1) we perform a further change
of
variables:
$$
\partial_+\widetilde E = {i\over 4\pi}\beta^{-1}\partial_+\beta \quad ,\quad
{\cal D}\widetilde E= \E^{-ic_V\Gamma[\beta]}{\cal D}\beta \quad .\eqno(A12)
$$
The partition function (2.1) is now obtained upon using the
Polyakov--Wiegmann identity ($A$2). When the partition function is written in
terms
of $\Sigma$, eq. ($A$11), we talk about the local formulation. In the form
(2.1) we say
that it is the non-local formulation, since formal integration over the
auxiliary field
$C$ leads to a non-local term in the $\beta$-fields.
\vskip .5cm
\noindent {\bf Acknowledgement}
\vskip .2cm

\noindent KDR would like to thank the CERN Theory Division for financial
support. EA and KDR further wish to thank for hospitality extended to them.

\vskip .5cm
\penalty-300
\centerline {\bf References}
\vskip .3cm\nobreak
\refer[[1]/E. Abdalla and M.C.B. Abdalla, hep-th/9503002,  to appear in Phys.
Rep.]

\refer[[2]/E. Abdalla and M.C.B. Abdalla, Int. J. Mod. Phys. {\bf A10} (1995)
1611]

\refer[[3]/E. Abdalla and K.D. Rothe,  Phys. Lett.  {\bf 363B} (1995) 85 ]

\refer[[4]/A. Dhar, G. Mandal and S. Wadia, Mod. Phys. Lett. {\bf A8} (1993)
3557;
D. Das, A. Dhar, G. Mandal and S. Wadia, Mod. Phys. Lett. {\bf A7} (1992) 71]

\refer[[5]/E. Abdalla and M.C.B. Abdalla, Phys. Lett. {\bf 337B} (1994) 347]

\refer[[6]/A.M. Polyakov and P.B. Wiegmann, Phys. Lett. {\bf 131B} (1983) 121;
{\bf
141B} (1984) 223; J. Wess and B. Zumino. Phys. Lett. {\bf 37B} (1971) 95; E.
Witten,
Commun. Math. Phys. {\bf 92} (1984) 455; K.D. Rothe, Nucl. Phys. {\bf B269}
(1986)
269]

\refer[[7]/D. Karabali and H.J. Schnitzer, Nucl. Phys. {\bf B329} (1990) 649]

\refer[[8]/D.C. Cabra, K.D. Rothe and F.A. Schaposnik,
 hep-th/9507043, to appear in Int. J. Mod. Phys. {\bf A}]

\refer[[9]/M. Blau and G. Thompson, Nucl. Phys. {\bf B408} (1993) 345]

\refer[[10]/O. Aharony, O. Ganor, J. Sonnenschein, S. Yankielowicz and N.
Sochen, Nucl. Phys. {\bf B399} (1993) 527]

\refer[[11]/E. Abdalla and M.C.B. Abdalla, hep-th/9503235,
to appear in Phys. Rev. {\bf D52} (1995) ]

\refer[[12]/M. Henneaux and C. Teitelboim, {\it Quantization of Gauge Systems}
(Princeton University Press, Cambridge, USA, 1992)]

\refer[[13]/E. Abdalla, M.C.B. Abdalla and K.D. Rothe,  {\it Non-Perturbative
Methods in Two-Dimensional Quantum Field Theory} (World Scientific,
Singapore, 1991)]

\refer[[14]/L.V. Belvedere, K.D. Rothe, B. Schroer and J. A. Swieca, Nucl.
Phys. {\bf B153} (1979) 112]

\refer[[15]/D. Gepner, Nucl. Phys. {\bf B252} (1985) 481]

\refer[[16]/Y. Frishman and J. Sonnenschein, Phys. Rep. {\bf 223} (1993) 309]

\refer[[17]/E. Abdalla and K.D. Rothe, Phys. Rev. {\bf D36} (1987) 3910]

\refer[[18]/H.J. Rothe and K.D. Rothe, Phys. Rev. {\bf D40} (1989) 545 ]

\refer[[19]/M. Wakimoto, Commun. Math. Phys. {\bf 104} (1989) 605]

\refer[[20]/V.G. Kac and D.A. Kazhdan, Adv. Math. {\bf 34} (1979) 97]

\refer[[21]/N. Gorsky and N. Nekrasov, Nucl. Phys. {\bf B414} (1994) 213]

\refer[[22]/I. Kogan and A. Zhitnitsky, preprint OUTP-95-25P, hep-ph/9509322]

\refer[[23]/D.J. Gross, I.R. Klebanov, A.V. Matytsin and A.V. Smilga,
Princeton preprint  PUPT-1577, hep-th/9511104.]

\end